\documentclass{Interspeech2024}
\usepackage{cite}
\usepackage{multirow}
\usepackage{pifont}
\usepackage{xcolor}
\usepackage{colortbl}
\usepackage{flushend}
\newcommand{\magenta}[1]{\textcolor{magenta}{#1}}

\interspeechcameraready

\title{Singing Voice Graph Modeling for SingFake Detection}

\name[affiliation={1}]{Xuanjun}{Chen}
\name[affiliation={1}]{Haibin}{Wu}
\name[affiliation={2}]{Jyh-Shing Roger}{Jang}
\name[affiliation={1}]{Hung-yi}{Lee}

\address{
  $^1$Graduate Institute of Communication Engineering, National Taiwan University\\
  $^2$Department of Computer Science and Information Engineering, National Taiwan University\thanks{Our code is available at: \magenta{https://github.com/xjchenGit/SingGraph.git}.}}
\email{\{d12942018, f07921092, hungyilee\}@ntu.edu.tw, jang@csie.ntu.edu.tw}


\begin{document}

\maketitle

\begin{abstract}
    
Detecting singing voice deepfakes, or SingFake, involves determining the authenticity and copyright of a singing voice. 
Existing models for speech deepfake detection have struggled to adapt to unseen attacks in this unique singing voice domain of human vocalization. 
To bridge the gap, we present a groundbreaking SingGraph model. The model synergizes the capabilities of the MERT acoustic music understanding model for pitch and rhythm analysis with the wav2vec2.0 model for linguistic analysis of lyrics.
Additionally, we advocate for using RawBoost and beat matching techniques grounded in music domain knowledge for singing voice augmentation, thereby enhancing SingFake detection performance.
Our proposed method achieves new state-of-the-art (SOTA) results within the SingFake dataset, surpassing the previous SOTA model across three distinct scenarios: it improves EER relatively for seen singers by 13.2\%, for unseen singers by 24.3\%, and unseen singers using different codecs by 37.1\%.
\end{abstract}

\section{Introduction}

The recent advancements in singing voice synthesis, exemplified by models such as VISinger \cite{zhang2022visinger} and DiffSinger \cite{liu2021diffsinger}, have brought significant progress to the field of singing voice generation. 
However, these developments have also raised concerns among artists, record companies, and publishing houses. 
The potential for unauthorized synthetic reproductions that closely imitate a singer's voice threatens the commercial value of the original artists. 
As society becomes increasingly aware of these issues, there is a pressing need to develop accurate methods for detecting deepfake singing voices. 
This is critical in safeguarding the rights and revenues of artists and ensuring the integrity of musical productions.

Since the singing voice is a form of human vocalization, it is logical to seek insights from the field of speech deepfake detection, commonly known as voice spoofing countermeasures (CM). 
Research in the CM domain has made significant progress, developing methods that effectively differentiate between genuine speech and spoofing attacks (fake speech) \cite{jung2022aasist, xue2023learning, ding2023samo, liao22_spsc}. 
Some of these breakthroughs have led to systems that achieve Equal Error Rates (EER) below 1\% on the ASVspoof2019 test set \cite{wang2020asvspoof}. 
However, despite these advances, there remains a challenge in generalizing these systems to unseen attacks, such as fake singing voices, and adapting to varied acoustic environments, as evidenced by their reduced effectiveness when confronted with real-world data \cite{muller2022does, liu2023asvspoof}.

To highlight the importance of improving the detection of fake singing voices, Zang et al. \cite{zang2023singfake} introduced the task of singing voice deepfake (SingFake) detection.
Singing voice deepfake detection presents unique challenges not found in speech deepfake detection. 
Unlike speech, singing involves adherence to melodies and rhythms that alter the pitch and duration of phonemes. 
Additionally, singing incorporates more varied artistic expressions and a broader spectrum of timbre, heavily influenced by the musical context. 
Furthermore, singing recordings often undergo significant editing and digital signal processing and are combined with instrumental accompaniments. 
Given these distinct characteristics of singing voices, the countermeasures designed for speech are not suited for directly applied to deepfake singing voice detection.
How to design a SingFake detection model tailored for the singing voice domain is still an open question.



In the paper, we introduce the SingGraph detection training framework for detecting singing voice deepfakes. 
SingGraph combines two advanced models: MERT \cite{li2023mert}, which understands music by focusing on pitch and rhythm, and wav2vec2 \cite{baevski2020wav2vec}, which captures phonemes in lyrics. 
This approach enables separate analyses of musical instruments and vocals,  we also demonstrate their complementary nature, as evidenced by the experiments with models A4-A6 in Table \ref{tab:ablation_comparison}.
Additionally, we are the first to introduce techniques like RawBoost \cite{tak2022rawboost} and beat matching, based on music domain knowledge, to enhance SingFake detection. 
Experiments on SingFake dataset reveal that SingGraph achieves 
state-of-the-art (SOTA) performance in various scenarios (i.e., the seen test, the unseen test, and the unseen codecs test).

\begin{figure*}[t]
\centering
\includegraphics[width=17cm]{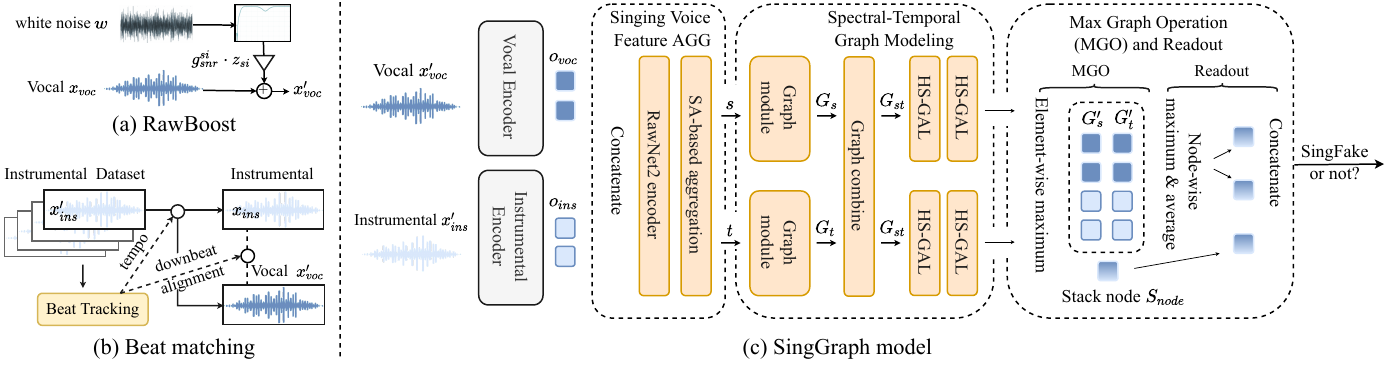}
\caption{The proposed SingGraph detection training framework.}
\label{fig:SingGraph}
\end{figure*}

\section{Related work}
\label{sec:speech_deepfake_detection}
In the SingFake dataset, Zang et al. \cite{zang2023singfake} develop four SingFake detection models based on four speech domain representative methods. 
First, AASIST \cite{jung2022aasist} utilizes the raw waveform as its feature input, employing graph neural networks alongside spectro-temporal attention mechanisms to analyze the audio data.
Next, \textbf{Spec-ResNet} is designed around the use of a linear spectrogram, which is obtained through a 512-point Fast Fourier Transform (FFT) with a hop size of 10 ms. The resulting spectrogram is then processed using the ResNet18 \cite{he2015deep} architecture, enabling the model to capture and analyze spectral features effectively.
\textbf{LFCC-ResNet} \cite{zhang2021one}, on the other hand, opts for Linear-Frequency Cepstral Coefficients (LFCC) as its speech feature. These 60-dimensional LFCCs are extracted from each audio frame, with the frame length set at 20ms and a hop size of 10ms, before being fed into the ResNet18 model. 
Lastly, \textbf{W2V2-AASIST} \cite{tak2022automatic} leverages the wav2vec2.0 model \cite{baevski2020wav2vec}, a self-supervised learning framework trained on extensive external speech datasets. This model is particularly adept at extracting nuanced phonetic and linguistic features from speech, providing a generalized representation.
However, none of the above methods consider capturing the musical information, such as pitch, rhythms, and so on.

\section{SingGraph detection training framework}


The SingGraph detection training framework, as shown in Figure~\ref{fig:SingGraph}, consists of two key components: singing voice augmentation techniques, including RawBoost and Beat matching, and the core SingGraph model. 
We use Demucs\footnote{https://github.com/facebookresearch/demucs} to separate a mixed singing song $x$ into instrumental $x_{ins}$ and vocal $x_{voc}$ components. 
Following separation, we apply data augmentation to create augmented instrumental $x'\_{ins}$ and vocal $x'\_{voc}$ tracks. 
The SingGraph model then processes these augmented inputs to verify whether the input is a genuine singing performance or a fake singing recording.

\subsection{Singing voice augmentation}
\label{lab:data_augmentation}

To boost performance, we leverage RawBoost for vocal augmentation and beat matching for instrumental augmentation.

\subsubsection{Beat matching for instrumental.}
In this part, we will introduce a data augmentation method based on music domain knowledge.
Musical compositions consist of various elements, including chord progressions, timbre, and beats, with beats being essential for setting the musical structure and alignment. 
In the audio domain, mixup \cite{zhang2017mixup} is a favored augmentation technique that blends pairs of audio tracks to expand the training dataset. 
However, mixing instrumental and vocal tracks with differing tempos can lead to a disordered and unattractive outcome. 
We propose the beat matching for instrumental augmentation to correspond to vocals as shown in Figure~\ref{fig:SingGraph}-(b). 
Firstly, we employ a cutting-edge beat tracking model ALL-in-one \cite{kim2023all} to determine each music track's tempo and downbeat map. 
We then organize each track into tempo-specific groups. 
During training, we randomly select an instrumental $x'_{ins}$ to replace the original pair instrumental $x_{ins}$ from the same tempo group to maintain consistent tempos. 
Additionally, we conduct downbeat alignment to choose an appropriate downbeat as the starting point for mixing singing voices. 
This pre-processing strategy enhances our ability to choose suitable music data for mixing, creating tracks that are harmoniously aligned in terms of tempo and downbeat. 

\subsubsection{RawBoost for vocal.}
We introduce a trick, inspired by stationary signal-independent additive noise with random coloration.
This trick significantly boosts speech deepfake detection by RawBoost\cite{tak2022rawboost, rosello2023conformer}.
Therefore, it is reasonable to adopt it for vocal augmentation in SingFake scenario, as the vocal part of singing also belongs to speech. 
As depicted in Figure~\ref{fig:SingGraph}-(a), the vocal input of the model, $x'{voc}$, is generated by the equation:
\begin{equation}
    x'_{voc} = x_{voc} + g^{si}_{snr} \cdot z_{si}
\end{equation}
where $z_{si}$ represents white noise $w$ that has been processed through a finite impulse response (FIR) filter, adjusted by a gain parameter $g^{si}_{snr}$ based on a random signal-to-noise ratio.

\subsection{SingGraph model}


This section introduces the proposed novel SingGraph model.
The SingGraph model, depicted in Figure~\ref{fig:SingGraph}-(c), integrates instrumental and vocal encoders based on self-supervised learning (SSL) models with a graph-based back-end model. 
This back-end leverages graph neural networks \cite{tak2021graph, velivckovic2017graph, jung2021graph}, which have led to significant advancements in various applications, including speaker verification \cite{jung2021graph} and spoofing detection \cite{tak2021graph, tak2021end, jung2022aasist}, by employing a structure of interconnected nodes and edges. 
It encompasses components for aggregating singing voice features, modeling spectral-temporal relationships, and executing readout operations, detailed in the following subsections.

\subsubsection{SSL-based instrumental and vocal encoder.}
\label{subsec:ssl_iv_encoder}

Speech SSL models like wav2vec2.0 \cite{baevski2020wav2vec, babu2021xls} excel in generalizing across downstream tasks \cite{yang2021superb}, where speech deepfake detection is one of applications \cite{wang2021investigating, tak2022automatic}, by capturing essential speech features, such as phonemes and tones, to distinguish deepfake speech. 
MERT \cite{li2023mert}, on the other hand, targets the musical aspects like rhythm and pitch through a multi-task mask language modeling approach, making it suited for music-related tasks \cite{du2023joint, yamamoto2023toward, yuan2024marble}, in the meantime, wav2vec 2.0's multi-lingual dataset training optimizes it for handling diverse singing voices.

MERT \cite{li2023mert} and wav2vec2 \cite{baevski2020wav2vec} are the instrumental backbone and vocal backbone models for instrumental and vocal encoders, respectively. 
Both instrumental and vocal encoders consist of a convolutional neural network (CNN) and a Transformer block. 
The instrumental and vocal encoder aims to model musical and vocal-related characteristics, respectively. 
The instrumental and vocal encoder will extract instrumental and vocal representation sequence $\{o_{ins}, o_{voc}\}$ from the input audio.
The experiments described in the paper utilized \textbf{MERT}\footnote{\href{https://huggingface.co/m-a-p/MERT-v1-330M}{huggingface.co/m-a-p/MERT-v1-330M}} and \textbf{wav2vec2.0 XLR-S}\footnote{\href{https://github.com/facebookresearch/fairseq/tree/main/examples/wav2vec}{github.com/facebookresearch/fairseq/tree/main/examples/wav2vec}} models, both initialized with pre-trained model from open-source repositories. 
Both models have 3 billion parameters.

\subsubsection{Singing voice feature aggregation.}
\label{subsec:singing_feature_agg}
The singing voice feature aggregation module leverages the RawNet2 encoder \cite{jung2020improved} and a self-attention-based (SA-based) aggregation layer \cite{tak2022automatic} to extract advanced spectral-temporal features, labeled as $\{s, t\}$. 
The RawNet2 encoder produces a detailed feature map, $\textbf{S}$, with dimensions in $\mathbb{R}^{C \times F \times T}$, indicating channels $C$, spectral bins $F$, and time frames $T$. 
Through the self-attention layer, this module effectively captures crucial spectral and temporal details. 
Specifically, the spectral feature $s$ is derived by aggregating across the spectral dimension $F$ of $\textbf{S}$, while the temporal feature $t$ is similarly extracted by summing over the time dimension $T$. 
This process ensures the module efficiently highlights relevant spectral and temporal information from singing voices.

\subsubsection{Spectral-temporal graph modeling.}
\label{subsec:st_graph_modeling}
The method employs the heterogeneous stacking graph attention layer (HS-GAL) \cite{wang2019heterogeneous} to analyze the varied aspects of singing voices, covering both instrumental and vocal components. 
Initially, it transforms spectral and temporal features $\{s, t\}$ into graphs $\{G_{st}, G_t\}$ through a graph module \cite{velivckovic2017graph}. 
The unified graph $G_{st}$ is generated by merging graphs through a graph combination module, employing a graph attention layer. 
This layer applies element-wise multiplication between nodes for symmetry, enabling attention weight calculations across spectral and temporal nodes, instead of concatenating nodes \cite{velivckovic2017graph}. 
This facilitates a detailed understanding of the intricate relationships in singing voice data. 
Moreover, an additional stack node is introduced to encapsulate the complex interactions between the spectral and temporal domains, linking to all nodes from $G_s$ and $G_t$. 
By applying HS-GAL layers in sequence, the stack node from the first layer is seamlessly integrated into the subsequent layer, enhancing the model's ability to capture and process heterogeneous relationships.
Stack nodes behave similarly to classification tokens \cite{devlin2018bert}.

\subsubsection{Max graph operation and readout.}
\label{subsec:readout_operation}
The final prediction module features a max graph operation (MGO) followed by a readout process. 
The MGO is designed to identify and integrate various spoofing-induced artifacts simultaneously, which is inspired by the element-wise maximum operations that are beneficial to detect anti-spoofing in several works \cite{tak2021end, lavrentyeva2017audio}.
It operates through two parallel branches, each applying an element-wise maximum on their outputs. Each branch consists of two sequential HS-GAL layers with a graph pooling layer following each HS-GAL, totaling four HS-GAL and four graph pooling layers within MGO. 
The HS-GAL layers within a branch share a stack node, passing it from one HS-GAL to the next, and an element-wise maximum is applied to the stack nodes from both branches. 
After element-wise maximum operations, we will obtain the spectral and temporal nodes $\{G'_s, G'_t\}$ of singing voice graphs and the stack node $S_{node}$.
Finally, a node-wise maximum and average are computed, and the last hidden layer is generated by concatenating the outcomes from these operations along with the stack node, ensuring a comprehensive representation of the data.

\begin{table*}[t]
\renewcommand{\tabcolsep}{5pt}
\renewcommand\arraystretch{1.3}
\small

\caption{Ablation study of SingGraph model for SingFake detection EER (\%).}
\vspace*{-0.2cm}
\centerline{
\begin{tabular}{c c c c|c c c c }
\hline
    & \textbf{Method} & \textbf{Augmentation} &  \textbf{Setup}  & \textbf{Train $\downarrow$} &  \textbf{T01 $\downarrow$} & \textbf{T02 $\downarrow$} & \textbf{T03 $\downarrow$}  \\

\hline
\hline
\multirow{2}{*}{A1)} & \multirow{2}{*}{W2V2-AASIST \cite{tak2022automatic}} & \ding{55} & $M$  &   \textbf{1.57}   &  \textbf{4.62}   &  \textbf{8.23}   &  13.62    \\
        &     & \ding{55} & $V$  &   1.70   & 5.39    &  9.10   &  \textbf{10.03}    \\
\hline
\multirow{2}{*}{A2)} & \multirow{2}{*}{W2V2-AASIST (Our)} & \ding{55} & $M$ &   2.24   &  5.31   &  15.80 &   16.37   \\
                 &     & \ding{55} & $V$  &   6.49   &  8.55   &  15.91 &   16.14      \\
\multirow{2}{*}{A3)} & \multirow{2}{*}{MERT-AASIST} & \ding{55}   & $M$ & 1.50 & 6.28 & 12.55 & 13.51  \\
                  & & \ding{55} & $V$  &  1.26 & 6.47 & 12.66 & 12.07  \\
\hline
\multirow{2}{*}{A4)} & \multirow{2}{*}{W2V2-AASIST} & RawBoost & $M$  &  1.78  &  5.47  &  9.90  &  9.66  \\
                  & & RawBoost & $V$  & 0.92 & 4.24 &  7.41 & 7.25  \\
\multirow{2}{*}{A5)} & \multirow{2}{*}{MERT-AASIST} & RawBoost   & $M$  & 0.88 & 5.54 & 12.66 & 12.79   \\
                  & & RawBoost & $V$  & 1.18 & 7.08 & 11.26 &  11.11  \\
\hline
\multirow{2}{*}{A6)} & \multirow{2}{*}{MERT-W2V2-AASIST} & RawBoost & $M$  &    1.20   &  4.31  &  9.79   &  8.85  \\
                     & & RawBoost & $V$  &   1.40  &   4.04   &  6.92   &   7.31  \\
\hline
\hline
\multirow{1}{*}{A7)} & \multirow{1}{*}{MERT-W2V2-AASIST}  & RawBoost & $IV$  &  0.84  &   \textbf{3.77}   &   6.84   &   7.05    \\
\hline
A8) & SingGraph  & RawBoost, Beat Matching & $IV$  &  \textbf{0.54} ({\color{red}+65.6})   &  4.01 ({\color{red}+13.2})  & \textbf{6.23} ({\color{red}+24.3}) & \textbf{6.30} ({\color{red}+37.1})  \\

\hline

\end{tabular}}

\label{tab:ablation_comparison}
\end{table*}

\section{Experiment}

\subsection{Experimental setup}
\label{subsec:setup}

\noindent\textbf{SingFake dataset:} 
The SingFake dataset \cite{zang2023singfake} offers 28.93 hours of genuine and 29.40 hours of deepfake song clips from popular user-generated platforms, organized into training, validation, and four test splits (T01 to T03). 
The training set features recordings from 12 singers, while the validation set introduces 4 new singers not included in the training. 
Test split T01 tests the model's recognition of Stefanie Sun, a singer from the training set. 
T02 and T03 challenge the model with 6 new singers, with T03 adding the twist of different communication codecs. 
Due to copyright concerns, the SingFake dataset has not been released. 
We've re-downloaded the necessary data from YouTube and Bilibili for experiments, but face download failures, leading to different data distribution compared to the original paper \footnote{Even though we failed to download some data, we still achieved the state-of-the-art. We will release the re-downloaded dataset metadata.}. 

\noindent\textbf{Model input setup:}
\cite{zang2023singfake} focused on comparing models using either mixed singing voice or vocal-only inputs. 
Our proposed SingGraph model, however, processes inputs divided into instrumental and vocal audio. 
We denote these input settings as ``$M$'' for mixture, ``$V$'' for vocals, and ``$IV$'' for separating instrumental and vocal audio and separately feeding them into the model.
Note that, in \cite{zang2023singfake}, they only consider ``$M$'' and ``$V$'', however, we propose ``$IV$'' for better modeling.

\noindent \textbf{Evaluation}: 
We will leverage the Equal Error Rate (EER) as our evaluation matrix (the same metric in \cite{zang2023singfake}), which is determined by setting a threshold on the produced scores where the false acceptance rate matches the false rejection rate. 

\subsection{Singing voice ablation study and analysis}

Table~\ref{tab:ablation_comparison} details our SingGraph model's ablation studies. 
A1 represents the results of the prior SOTA collected from the SingFake paper \cite{zang2023singfake}. 
Due to music data copyright issues as described in Section~\ref{subsec:setup}, we failed to download some data, and we reproduced the W2V2-AASIST model, labeled as model A2, for a fair comparison. 
Models A3 and A5 feature the MERT-AASIST architecture, replacing W2V2-AASIST's W2V2 encoder with MERT. 
Models A6-A8 follow the architecture in Figure~\ref{fig:SingGraph}-(c) but vary in augmentation and input settings.

\subsubsection{Preliminary ablation study}
In our preliminary study, focusing on models A2 to A6 in Table~\ref{tab:ablation_comparison}, we evaluate the efficacy of RawBoost and the synergistic potential between MERT and W2V2.
\textbf{(1) RawBoost.} 
After analyzing models A2 and A4, we observe significant improvements in W2V2-AASIST from T01 to T03 for both mixture and vocal inputs after adding RawBoost. 
However, for MERT-AASIST (A3 and A5), adding RawBoost showed fluctuations without clear improvements.
\textbf{(2) The complementary of MERT and wav2vec2.0.}
We assess the synergy between W2V2 and MERT by comparing rows A4, A5, and A6. Generally, model A6 (MERT-W2V2-AASIST) outperforms both MERT-AASIST and W2V2-AASIST models.

Experiment findings indicate RawBoost significantly improves wav2vec2.0's extraction of vocal features, yet it barely improves MERT's extraction of music-related features. 
Additionally, the results suggest that using MERT for music information and wav2vec2.0 for vocal features can complement each other with better results. 
So, in the subsequent experiments, RawBoost only augments the vocal part.

\subsubsection{SingGraph model ablation study.}

In our model ablation study, examining models A6 to A8 in Table~\ref{tab:ablation_comparison}, we aim to assess the necessity of each SingGraph model component's design.
\textbf{(1) Instrument and vocals.}
Analyzing models A6 and A7 reveals that vocal-only inputs enhance the MERT-W2V2-AASIST performance over mixed inputs across test sets T01 to T03. 
Model A7, using separate instrumentals for MERT and vocals for W2V2, outperforms its counterpart in A6 with significant margins: improvements of 0.56\%, 0.27\%, 0.08\%, 0.26\%, and 1.51\% are noted across the training and test sets T01 to T03, respectively.
\textbf{(2) RawBoost and beat matching.}
The primary difference between models A7 and A8 is that A8 uses beat-matching augmentation to instrumentals. 
This augmentation significantly enhances performance, achieving the lowest EER of 0.54\% on the training set. 
Moreover, A8 excels in test sets T02 and T03.
Its performance on test set T01 is also comparable with A7.

The experiments highlight the effectiveness of using specialized encoders for vocal and instrumental parts to capture the intricacies of singing voices from different perspectives. 
RawBoost and Beat Matching augmentations enhance vocal and instrumental features, making them a powerful combo. 
In SingFake detection, this approach optimizes the analysis of voice nuances and instrumental details, reducing EER. 
By focusing on vocal authenticity and the harmony of instrumental sounds, the strategy effectively reduces errors, enabling the precise identification of real versus fake singing clips.

\begin{table}[t]
\renewcommand{\tabcolsep}{2pt}
\renewcommand\arraystretch{1.3}
\small

\caption{Comparison of SOTA SingFake detection EER (\%).}
\vspace*{-0.2cm}
\centerline{
\begin{tabular}{cc|c c c c}
\hline
  \textbf{Method} &  \textbf{Setup}  & \textbf{Train $\downarrow$}   &  \textbf{T01 $\downarrow$}  & \textbf{T02 $\downarrow$}  &  \textbf{T03 $\downarrow$}  \\

\hline
\hline
AASIST \cite{jung2022aasist} &  $M$  & 4.10 & 7.29 & 11.54 & 17.29  \\
             &  $V$  &   3.39   &   8.37  &   10.65 &   13.07  \\
Spec-ResNet \cite{he2015deep}  &  $M$  &   4.97   &  14.88  &  22.59  &  24.15  \\
             &  $V$  &   5.31   &  11.86  &  19.69  &  21.54   \\
LFCC-ResNet \cite{zhang2021one}  &  $M$  &   10.55  &  21.35  &  32.40  &  31.85   \\
             &  $V$  &   2.90   &  15.88  &  22.56  &  23.62   \\
W2V2-AASIST \cite{tak2022automatic} &  $M$  &   \textbf{1.57}   &   \textbf{4.62}  &  \textbf{8.23}   &  13.62   \\
             &  $V$  &   1.70   &   5.39  &  9.10   &  \textbf{10.03}  \\
\hline
\multirow{2}{*}{SingGraph}    &  \multirow{2}{*}{$IV$}  &  \cellcolor{lightgray!10}{\textbf{0.54}}   &   \cellcolor{lightgray!10}{\textbf{4.01}}  &  \cellcolor{lightgray!10}{\textbf{6.23}}   &  \cellcolor{lightgray!10}{\textbf{6.30}} \\
             &  & ({\color{red}+65.6}) & ({\color{red}+13.2}) & ({\color{red}+24.3}) & ({\color{red}+37.1}) \\
\hline
\end{tabular}
}

\label{tab:sota_comparison}
\end{table}

\subsection{Comparison with SOTA SingFake detection model}

In Table~\ref{tab:sota_comparison}, we benchmark SingGraph model against former leading models such as AASIST \cite{jung2022aasist}, Spec-ResNet \cite{he2015deep}, LFCC-ResNet \cite{zhang2021one}, and W2V2-AASIST \cite{babu2021xls}, as discussed in Section \ref{sec:speech_deepfake_detection}. 
Our SingGraph outperforms the previous SOTA model, W2V2-AASIST, demonstrating significant relative improvements: 65.6\% in training, and 13.2\%, 24.3\%, and 37.1\% in the T01, T02, and T03 test sets, respectively. 
These results underscore SingGraph's robust capability in accurately detecting both seen and unseen SingFake instances across various codecs, showcasing its effective generalization in SingFake detection.

However, Our SingGraph model struggles with the Persian test set, notably influenced by hip-hop, contrasting with the rock and ballad genres in other sets. This points to a limitation in its ability to generalize across diverse musical genres, signaling an area for future improvement.

\section{Conclusion}

This paper proposes a novel SingGraph framework, which combines MERT and wav2vec2.0 to analyze vocal and musical nuances, to address singing voice deepfake. 
The introduced RawBoost and beat matching augmentations significantly improve detection performance. 
SingGraph achieves new SOTA results on the SingFake dataset, relatively improving EER for seen singers by 13.2\%, unseen singers by 24.3\%, and unseen singers across different codecs by 37.1\%.




\section{Acknowledgements}
This work was partially supported by the National Science and Technology Council, Taiwan (Grant no. NSTC 112-2634-F-002-005, Advanced Technologies for Designing Trustable AI Services). We also thank the National Center for High-performance Computing (NCHC) of National Applied Research Laboratories (NARLabs) in Taiwan for providing computational and storage resources.

\bibliographystyle{IEEEtran}
\bibliography{mybib}

\end{document}